
\input harvmac
\def\Tr{\hbox{\rm Tr}}
\def\diag{\hbox{\rm diag}}
\def\weff{W_{\rm\scriptstyle eff}}
\def\psbar{\overline{\psi}}
\def\to{\rightarrow}
\def\Pibar{\overline{\Pi}}
\def\pint{\int \kern-1em -}
\Title{\vbox{\baselineskip12pt\hbox{FERMILAB--PUB--92/35--T}
\hbox{hepth@xxx/9202050}}}
{\vbox{\centerline{The Marinari-Parisi Model and}
\medskip
\centerline{Collective Field Theory}} }
\centerline{J.D.~Cohn and H.~Dykstra\footnote{$^\dagger$}
{bitnet jdcohn@fnal and dykstra@fnal}}
\bigskip
\centerline{Fermi National Accelerator Laboratory}
\centerline{P. O. Box 500, Batavia, IL  60510}
\bigskip
\centerline{\bf Abstract}
\smallskip
We derive the supersymmetric collective field theory for the
Marinari-Parisi model.
For a specific choice of the superpotential, to leading order
we find a one parameter family of
ground states which can be connected via instantons.
At this level of analysis the instanton size implied by the underlying matrix
model does not appear.
\Date{2/11/92}  
\lref\rev{See, for example,
J.~Louis, ``Status of Supersymmetry Breaking in String Theory'',
SLAC-PUB-5645, talk at Particles and Fields '91 Symposium.}
\lref\flow{M.~Douglas, N.~Seiberg and S.~H.~Shenker,
Phys. Lett. B244 (1990), 381.}
\lref\Dab{A.~Dabholkar, ``Fermions and Nonperturbative Supersymmetry
Breaking in the One Dimensional Superstring'', Rutgers preprint RU-91-20
(May 1991).}
\lref\bipz{E. Brezin, C. Itzykson, G. Parisi and J.B. Zuber, Comm. Math.
Phys. 59 (1978) 35.}
\lref\pg{P.~Ginsparg and J.~Zinn-Justin, Phys. Lett. B240 (1990), 333.}
\lref\dg{U.~Danielsson and D.~Gross, Nucl. Phys. B366 (1991), 3.}
\lref\dan{U.~Danielsson, ``Symmetries and Special States in
Two Dimensional String Theory'', Princeton preprint PUPT-1301 (1991).}
\lref\DJR{K.~Demeterfi, A.~Jevicki, J.P.~Rodrigues, Nucl. Phys. B362 (1991),
173; Nucl. Phys. B365 (1991), 499\semi
I.~Klebanov, ``String Theory in Two Dimensions'' (Trieste
Lectures), Princeton Preprint PUPT-1271 (1991).}
\lref\greg{G. ~Moore and N.~Seiberg, ``From Loops to Fields in 2D
Quantum Gravity'', Rutgers and Yale Preprint RU-91-29/YCTP-P19-91(1991).}
\lref\aj{J.~Avan and A.~Jevicki, ``Classical Integrability and Higher
Symmetries of Collective String Field Theory'', Brown Preprint
Brown-HET-801 (1991).}
\lref\witten{E.~Witten, ``Ground Ring of Two Dimensional String Theory'',
IAS Preprint IASSNS-HEP-91/51 (1991).}
\lref\pol{J.~Polchinski, ``Classical Limit of 1+1 Dimensional String Theory'',
Univ. of Texas preprint UTTG-06-91 (1991).}
\lref\poltachy{J. ~Polchinski, Nucl. Phys. B346 (1990), 253.}
\lref\mpy{D.~Minic, J.~Polchinski, Z.~Yang, Univ. of Texas preprint
UTTG-16-91.}
\lref\twomat{T.~Tada and M.~Yamaguchi, Phys. Lett. B250 (1990), 38\semi
E.~Martinec, Comm. Math. Phys. 138 (1990), 437\semi
M.~Douglas, Rutgers and Ecole Normale preprint(1990).}
\lref\vip{J.~D.~Cohn and V.~Periwal, Phys. Lett. B270 (1991), 18.}
\lref\marte{M.~Martellini, M.~Spreafico, K.~Yoshida, ``A Continuum String
for $D>1$'', University of Rome preprint INFN-853(1992).}
\lref\gregpl{G.~Moore, M.R.~Plesser, S.~Ramgoolam, Yale preprint
YCTP-P35-91 (1991).}
\lref\Jnp{A.~Jevicki, ``Nonperturbative Collective Field Theory'', Brown
preprint BROWN-HET-807 (1991).}
\lref\GK{D.~Gross and N.~Miljkovic, Phys. Lett. 238B (1990), 217;
E.~Brezin, V.~Kazakov and Al.~B.~Zamolodchikov, Nucl. Phys. B338 (1990), 673;
G.~Parisi, Phys. Lett. 238B (1990), 209.}
\lref\gpar{G.~Parisi, talk at Cargese Workshop, 1990.}
\lref\MP{E.~Marinari and G.~Parisi, Phys. Lett. B240 (1990), 375.}
\lref\MK{M.~Karliner and A.~Migdal, Mod. Phys. Lett. A5 (1990), 2565.}
\lref\DJ{S.~Das and A.~Jevicki, Mod. Phys. Lett. A5 (1990), 1639.}
\lref\Sak{A.~Jevicki and B.~Sakita, Nucl. Phys. B165 (1980), 511.}
\lref\Sakbook{B.~Sakita, {\it Quantum Theory of Many-Variable Systems and
Fields}, World Scientific (1985).}
\lref\Jsup{A.~Jevicki, J.~Rodrigues, ``Supersymmetric
Collective Field Theory'', Brown preprint BROWN-HET-813 (1991).}
\lref\Dav{F. David, Nucl. Phys. B348 (1991), 507.}
\lref\lecht{A.~Jevicki, Nucl. Phys. B146 (1978), 77 \semi
O.~Lechtenfeld, ``Semiclassical Approach to Finite-N Matrix
Models'', IAS preprint IASSNS--HEP--91/86.}
\lref\shs{S.~H.~Shenker, Rutgers preprint RU-90-47, to appear in
proceedings of Cargese 1990 Workshop.}
\lref\mm{E.~Brezin and V.~Kazakov, Phys.Lett. B236 (1990), 144\semi
M.~Douglas and S.~H.~Shenker, Nucl. Phys. B335 (1990), 635\semi
D.~Gross and A.~Migdal, Phys. Rev. Lett. 64 (1990), 127.}
\lref\shahar{S. Ben-Menahem, ``D=0 Matrix Model as Conjugate Field Theory,
SLAC preprint 5377(1992, preliminary version Nov 1990)}.
\lref\subl{This subleading term has appeared in other contexts,
for example in \shahar\semi
O. Lechtenfeld, ``On Eigenvalue Tunneling in Matrix Models'', IAS preprint
IASSNS-HEP-91/2(Jan. 1991)
\semi the preliminary version of \DJ (Spring 1990) \semi
D. Karabali and B. Sakita, Int.J.Mod.Phys.A6 (91) 5079 \semi
J.D.Cohn and S.P. deAlwis, ``String Field Theory for d$<=$1 Matrix Models,''
Colorado and IAS preprint COLO-HEP-247/IASSNS-HEP-91/7(Feb. 1991), to appear
in Nucl. Phys. B and its importance has been emphasized by S. Wadia, fall 1990,
as well as in \refs{\lecht,\Jnp}.}
\newsec{Introduction}
Nonperturbative effects are not yet understood
in string theory.  Since much important string physics relies upon these \rev,
it is important to understand any known examples.
In the matrix models descriptions of non-critical strings \mm, a source of
both supersymmetry breaking and other nonperturbative effects
is one eigenvalue tunneling processes \refs{\flow,\MP,\shs,\gpar}.
One string theory which exhibits supersymmetry breaking nonperturbatively
is the Marinari-Parisi model \MP.  In an attempt to understand better the
nature of the nonperturbative physics found there, in this paper
we transform the model to collective
fields.  This transformation led to much insight about the spacetime
interpretation of the $d=1$ bosonic model \DJ.  We then consider the
supersymmetry breaking seen previously in the matrix
description \refs{\MP,\MK,\gpar,\Dab} and close with some comments about
the current status of the spacetime identification of the model.
\newsec{The Marinari-Parisi Supermatrix Model}
The one-dimensional string is described by a two-dimensional worldsheet
embedded in one spacetime dimension.  This may be approximated by a
triangulated surface with an additional degree of freedom on the faces of the
triangulation which describes its position in the one-dimensional space.  This
leads to the matrix model description of the $d$=1 string which has a single
matrix function of one spacetime variable \refs{\GK,\pg}.

The one-dimensional superstring has worldsheet supersymmetry, which leads to
supersymmetry in the spacetime spectrum of the superstring.  We do not know how
to build a matrix model which describes a theory with worldsheet supersymmetry,
but we can impose spacetime supersymmetry by describing surfaces imbedded in
one-dimensional superspace.  This is the Marinari-Parisi \MP\
model for one-dimensional superstrings.  The action of
this model is:
\eqn\ssaction{
S = N \int dt\,d\bar{\theta}\,d\theta\,  \Tr \bigl[\half\bar{D}\Phi \, D\Phi
              + W(\Phi)\bigr],
}
where $D$ is the differential operator on superspace and $\Phi$ is a
hermitian $N\times N$ matrix-valued superfield.  In components, the expansion
of $\Phi$ is
\eqn\comp{
\Phi = M + \bar{\theta}\Psi + \bar{\Psi}\theta + \bar{\theta}\theta F}

The matrix superfield $\Phi$ cannot be diagonalized by a unitary rotation.
However, there exists a consistent truncation to a supersymmetric subsector of
the Hilbert space, where $M$ is diagonalized \Dab.  To define this truncation
let $U$ be the unitary
matrix such that $UMU^{\dagger}= \diag(\lambda_i)$.  Then we restrict our
theory to only those states generated by the diagonal elements $\psi_i =
(U\Psi U^{\dagger})_{ii}$ acting on the vacuum, whose wavefunctions
depend only on the eigenvalues $\lambda_i$.  This theory is described by an
action with $N$ superfields $X_i$ and an effective superpotential which
incorporates the Jacobian for this change of variables.
\eqn\traction{
S = N \int dt\,d\bar{\theta}\,d\theta \left(\sum_i \bigl[\half\bar{D}X_i \,
                DX_i \bigr] + \weff (X) \right)
}
\eqn\effspot{
\weff (X) = \sum_i W(X_i) -
      {1\over N} \sum_{i < j} \ln (X_i - X_j)
}
The component field expression for the superfields is $X_i = \lambda_i +
\bar{\theta} \psi_i + \bar{\psi}_i \theta + \bar{\theta}\theta f_i$.  In terms
of the components the supercharge of this theory is
\eqn\scharge{
Q = -{i\over N} \sum_i \left( {\partial\over \partial\lambda_i} -
            N {\partial\weff(\lambda)\over \partial\lambda_i} \right) \psi_i
}
and the Hamiltonian,
\eqn\ham{
H = \ha \sum_i \left( -{1\over N^2}{\partial^2\over \partial\lambda_i{}^2 }
       + \left|{\partial\weff\over \partial\lambda_i}\right|^2
       - {1\over N}{\partial^2\weff\over\partial\lambda_i{}^2} \right)
       + {1\over N}\sum_{i,j} \psi_i^* {\partial^2\weff\over
            \partial\lambda_i\partial\lambda_j} \psi_j.
}
This theory was considered at length in the eigenvalue description in
ref.~\Dab.

What is the interpretation of this truncation?  In the bosonic $c$=1 matrix
model the dynamics of the eigenvalues describes the singlet sector of the
theory, that is, operators such as $\Tr M^n$ which do not depend on the angular
variables $U_{ij}$.
These operators may be generalized by
replacing the matrix $M$ by the superfield $\Phi$.  The components
of these operators such as $\Tr (\Psi M^n)$ or $\Tr (\bar{\Psi} M^m \Psi M^n)$
act within the diagonal sector (of $M$) of the theory.  So this truncation
is a consistent supersymmetric counterpart to the truncation
to the
eigenvalue variables in the bosonic case.  Since the supercharge for the
full theory does not take states out of the truncated sector,
the calculation of quantities such as
$\langle\hbox{\rm anything}|Q|\hbox{\rm state in truncated sector}\rangle$, the
trademark of supersymmetry breaking when
$|\hbox{\rm state in the truncated sector}\rangle$ is the vacuum, are valid for
the theory as a whole.

\newsec{Supersymmetric Collective Field Theory}
We would like to treat this theory using the collective field method
\refs{\DJ,\Sak}.  To begin introduce the density variables for the
eigenvalues $\lambda_i$:
\eqn\density{
\phi_k = \sum_i e^{ik\lambda_i}.
}
Only $N$ of these variables are independent.  In the $N\to\infty$ limit these
become the Fourier modes of the density $\phi(x) = \sum_i \delta(x-\lambda_i)$,
with the constraint $\int \phi = N$.
This is the usual collective field for the bosonic $d$$=$$1$ theory.  To
complete the field content of this theory, introduce fermionic fields:
\eqn\fermions{
\eqalign{&\psi_k = \sum_i \psi_i e^{ik\lambda_i}          \cr
         &\psbar_k = \sum_i \bar{\psi}_i e^{ik\lambda_i} \cr}
}
In the large N limit these variables become the Fourier components of fermionic
partners to the bosonic collective field.

To quantize this theory we introduce canonical momenta $p_i,\pi_i,\bar{\pi}_i$
for the eigenvalue variables $\lambda_i,\psi_i,\bar{\psi}_i$ and similarly
$p_k,\Pi_k,\Pibar_k$ conjugate to $\phi_k,\psi_k,\psbar_k$, with
Poisson brackets
\eqn\pbrak{
\eqalign{&\{p_k,\phi_q\} = \delta(k+q), \cr
         &\{\Pi_k,\psi_q\} = \delta(k+q), \cr
         &\{\Pibar_k,\psbar_q\} = \delta(k+q), \cr
         &\hbox{\rm (all others zero).} \cr}
}
In addition there are constraints corresponding to the fermionic momenta.  In
the eigenvalue variables these are determined by varying the action
\traction\ with respect to $\dot{\psi}_i$:
\eqn\constr{
\eqalign{&\chi_i = \pi_i - {\textstyle i\over 2}\bar{\psi}_i = 0, \cr
         &\bar{\chi}_i = \bar{\pi}_i + {\textstyle i\over 2} \psi_i = 0. \cr}
}
By using the canonical change of variables these may be rewritten in terms of
the density variables:
\eqn\collcon{
\eqalign{&\chi_k = \sum_i \e{ik\lambda_i} \chi_i
                 = \phi_{k+q} \Pi_{-q} - {\textstyle i\over 2} \psbar_k \cr
   &\bar{\chi}_k = \sum_i \e{ik\lambda_i} \bar{\chi}_i
                 = \phi_{k+q}\Pibar_{-q} + {\textstyle i\over 2} \psi_k \cr}
}
These constraints can be formally solved to give
$\Pi_k = {i\over 2}(\raise.5ex\hbox{$\psbar$}\kern-.2em
/\kern-.1em\lower.3ex\hbox{$\phi$})_k$.  With the
constraints we use the Dirac quantization procedure to find the commutation
relations of the density variables:
\eqn\commutator{
\eqalign{ [p_k,\phi_q] = -i\delta(k+q) \cr
          \{\psbar_k,\psi_q\} = \phi_{k+q} \cr
          [p_k,\psi_q] = \Pibar_{k+q}. \cr}
}
In the large N limit these become the commutators of continuous fields
$\phi(x), \psi(x)$.  These commutators agree with those found by Jevicki and
Rodrigues \Jsup\ by supersymmetrizing the bosonic $d$$=$$1$ collective field
theory.

With the quantization complete we rewrite the Hamiltonian in
terms of the new variables.  By the canonical change of variables we have an
expression for $p_i$:
\eqn\momrel{
p_i = {\partial\phi_k\over\partial\lambda_i} p_{-k}
      + {\partial\psbar_k\over\partial\lambda_i} \Pibar_{-k}
      + \Pi_{-k} {\partial\psi_k\over\partial\lambda_i}.
}
This is a classical expression for the relationship between the canonical
variables.  After quantization the variables become operators with non-trivial
commutation relations.  The classical expression does not fix the ordering of
these operators, but the expression given for the fermionic part
is the only one consistent with
the requirements that $p_i$ be a Hermitian operator and $p_i \left| 0
\right\rangle = 0$.

Inserting this expression into \scharge\ gives the collective field
supercharge and the anti-commutator $H = {1\over 2}\{Q,\bar{Q}\}$
is the Hamiltonian.
We need the quantities $\sum_i p_i \psi_i$ (for $Q$) and
$\sum_i \bar{\psi}_i p_i$ (for $\bar{Q}$); using \momrel\ it may be seen that
$Q$ and $\bar{Q}$ are not naively Hermitian conjugates of each other in
collective field variables, but $Q$ has an additional term:
\eqn\jac{
\sum_{k} ik [{\rm e}^{ik\lambda_i}, p_{-k}] \psi_i
}
which can be traced back to the fact that $p_k^\dagger \ne p_{-k}$ because of
the Jacobian for the change to collective variables.  However, the commutator
can be calculated in the purely bosonic theory, where it is already known
that the similarity transform which restores the naive Hermiticity properties
is \Sak:
\eqn\simtran{
\partial p \to \partial p - \half i{\partial\phi\over\phi}.
}
All other fields remain unchanged under this transformation.  This can be
understood from the fact that the change of variables is linear in the
fermionic degrees of freedom, hence the Jacobian depends only on $\phi$, which
commutes with all fields except $p$.

The transformed supercharge in collective field variables is:
\eqn\collQ{
Q = \int dx \, {1\over N}\partial\sigma(x)\psi(x) + i\left(
      W'(x) - {1\over N}\int {dy \, \phi(y) \over x-y}
          + {1\over 2N} {\phi'(x) \over \phi(x)} \right) \psi(x),
}
and $\bar{Q}$ is the naive conjugate.  The field $\sigma$ has been defined by
\eqn\sig{
\partial\sigma = \partial p - {i\over 2} {\psbar\over\phi} \partial \!\left(
                   {\psi\over\phi} \right) + {i\over 2} \partial \!\left(
                   {\psbar\over\phi} \right) {\psi\over\phi}
}
We note that in effect the superpotential has picked up a new term from the
collective field Jacobian\subl.  The final result for the Hamiltonian is
\eqn\hamcoll{
\eqalign{
H = \int dx\; \left[ {1\over 2N^2}\right. &
                     \partial\sigma(x)\phi(x)\partial\sigma(x) +
        \ha \phi(x)\left( W'(x) - {1\over N}\pint dy {\phi(y)\over x-y}
                     + {1\over 2N} {\phi'(x)\over\phi(x)} \right)^2 \cr
     {}  &  + \left. {1\over N} W''(x){\psbar(x)\psi(x)\over\phi(x)} +
    {1\over N^2}\psbar(x)\partial_x\pint dy {\psi(y)\over (x-y)} \right]. \cr}
}

A term proportional to
$ \int dx [\partial \sigma(x), \psi(x)]$ has been
set to zero (its value in one regularization scheme) and the constraint
$\int dx \, \phi(x) = N$ is implicit.
For a generic choice of superpotential this Hamiltonian is apparently nonlocal
due to the instantaneous interaction between the eigenvalues.  However, for the
special case of a cubic superpotential the nonlocality vanishes to leading
order in $1/N$.  In
particular we may choose $W(x) = \ha(gx - {1\over 3} x^3)$ as considered in
\Dab.  Then the bosonic potential terms of the theory can
be seen to be equivalent to (dropping the subleading $\phi'/\phi$ term):
\eqn\local{
\eqalign{
V &= {1\over 2} \int dx \,\left[ W'(x)^2 \phi(x)+
{\pi^2 \over 3N^2} \phi(x)^3 -
     {1\over N}  \pint dy\,{W'(x)-W'(y) \over x-y } \phi(x) \phi(y) \right] \cr
  &\hbox{\rm (for specific $x^3$ potential)}
= {1 \over 2}\int dx \, \left[ ( W'(x)^2 - x ) \phi(x) + {\pi^2 \over 3N^2}
\phi(x)^3 \right], \cr  }
}
making use of the constraint $\int \phi = N$ and using the assumption that
the support of $\phi$ is nonsingular to obtain $(\pi^2/3)\int dx \phi^3(x)$
from
$\int dy \phi(y) (\pint dx \phi(x)/(x-y))^2$ \refs{\Sak,\Sakbook}.

\newsec{Ground States of the Collective Field Theory}
To investigate this theory further we consider the ground state.
As usual in a supersymmetric theory, this is expected to be a static field
configuration with zero potential energy in the zero fermion number sector.
A look at eqn.~\hamcoll\ shows us
that we must solve the equation
\eqn\dodo{
W'(x) - {1\over N} \pint dy {\phi(y) \over x-y}
+ {1\over 2N} {\phi'(x) \over \phi(x)} = 0
}
This equation may be formally integrated to give
$\phi \propto {\rm exp}(-2N\weff(\phi;x))$.  However, this result may be
misleading for several reasons.  In the particular case of the cubic
superpotential above this is inconsistent with the normalization constraint
on $\phi$.  That the ground state is not normalizable might be interpreted as
a signal of supersymmetry breaking, but there is another difficulty as well.
At large values of $x$ we expect $\phi(x)$ to be exponentially
small; but the term in \dodo\ arising from the collective field Jacobian is
just the first term in a series which is particularly badly behaved when
$\phi$ is small \lecht.  So the equation can only be trusted in the region
where $\phi$ is large. In this case the Jacobian term is down by $1/N$ and
can be neglected at the sphere level.

Thus we are led to consider an equation
which is formally identical to the BIPZ \bipz\ method
for bosonic matrix models,
where now the superpotential plays the role of the potential.  We may expect
that the critical points of these models may be classified by the critical
behavior of this equation; i.e., eigenvalues spilling over the barriers in the
{\it superpotential}.  However, the interpretation will be different, since
both minima and maxima of the superpotential correspond to local minima of
the potential.  Thus, for instance, in the cubic superpotential criticality
occurs when the eigenvalue density reaches the top of a quadratic maximum of
the
superpotential; in spacetime we see instead a coalescing of a second potential
well with the endpoint of the eigenvalue density to form a {\it cubic} critical
point.  There does not appear to be a critical point which is quadratic in the
spacetime potential (as for the bosonic $c$=1 theories).  As long
as supersymmetry is unbroken, solutions to an equation of the form \dodo\
corresponding to a one matrix model configuration appear to rule out $-x^2$
critical behavior in the actual potential.  One way to see this is to start
with the $-x^2$ potential and work backwards to the corresponding
superpotential.
The eigenvalue density on two sides of a $-x^2$ critical
point in the potential corresponds to eigenvalue density in the minimum and
maximum of the corresponding superpotential, with one of the densities
negative.
Higher order potentials
will also induce explicit nonlocal interactions between the
eigenvalues in the bosonic sector; in some cases
this may become local for symmetry reasons \Jsup\ or disappear in the
double scaling limit.

For the cubic superpotential $W=gx-{1\over 3}x^3$ this equation has been solved
in \Dab:
\eqn\groundzero{
\eqalign{
{}& \phi_0 =  -{N\over 2\pi} (x+a+b) \sqrt{(2a-x)(x-2b)},
              \quad 2b\le x \le 2a, \cr
{}& g - (a+b)^2 - \half (a-b)^2 = 0, \cr
{}& 2 + (a-b)^2(a+b) = 0. \cr}
}
The potential for fluctuations around this background,
away from the support of $\phi_0$, is linear in the fluctuation (since here
the fluctuation is constrained $\delta\phi\ge 0$).
The lowest order term is
\eqn\potent{
{\textstyle{1\over 8}} (x+a+b)^2 (x-2a)(x-2b) \,\delta\phi(x), \quad x<2b,
x>2a.
}
The potential is zero at $x=-(a+b)$.  In \refs{\Dav,\Dab}
this was interpreted as an
alternate classical ground state for the highest eigenvalue.  In supersymmetric
quantum mechanics the presence of two ground states signals the possibility of
supersymmetry breaking by instanton tunneling from one well to another.

In the collective field theory this extra ground state is represented by a
singular field configuration with $\delta$-function support at $x\sim -(a+b)$:
\eqn\grsttwo{
\phi \sim \phi_c + \alpha\,\delta(x+a+b).
}
In this equation $\phi_c$ represents a continuous distribution, equal to
$\phi_0$ in \groundzero\ to leading order in $1/N$, and the parameter
$\alpha$ in the matrix model picture
counts the number of eigenvalues sitting in the second
well.  In the collective field theory a solution of this form may be found
perturbatively in $\alpha/N$
by a simple modification of the BIPZ procedure.
The double scaling limit of \Dab\ requires $\alpha$ to be finite (or zero) as
$N\to\infty$.

In the eigenvalue picture of \Dab, it is clear that there are
precisely two
ground states, corresponding to eqn.~\groundzero\ and eqn.~\grsttwo\ with
$\alpha$=$1$, that is, one eigenvalue sitting near $x=-(a+b)$.
The repulsion of eigenvalues prevents more than one eigenvalue from living
at this point.  It is easily verified that with an ansatz of the form
\grsttwo\ there are no further zeros of the potential for real values of
the eigenvalues, so there are no additional ground states.  The existence of
two degenerate ground states with zero energy perturbatively implies that
supersymmetry is broken, and nonperturbative contributions give a
non-zero ground state energy as calculated in \Dab.

By contrast, the field theory seems to allow a
continuum of ground states assuming the perturbative
expansion in $\alpha/N$ has a finite radius of convergence.  These involve
singular field configurations which cannot be simply regarded as a limit of
smooth field configurations.  (The self interactions of a `$\delta$-function'
of finite width are self-repulsive.  As a result, configurations
energetically prefer to spread out rather than approach a singular
$\delta$-function, unless the principal value prescription is interpreted more
broadly.)  Further, these singular field configurations give singular
contributions from the collective field Jacobian which is naively down
by $1/N$ (although it is expected that the singular higher order in $1/N$
terms in the Jacobian will be important in this
case).  Since the neglected terms from the Jacobian are non-linear in $\phi$,
they could either fix $\alpha$ or destroy the solution
entirely.  Neglecting the superpotential in \hamcoll, the bosonic sector
is the theory considered by Jevicki \Jnp. He found a soliton solution with
fixed coefficient corresponding to single eigenvalue motions.
{}From the underlying matrix model we might speculate this is
true in our case also.  We have not
found a transformation analogous to the one in \aj\ which takes the
potential ($-x^2$ in that case) to zero and it is not clear how to
consistently treat
the subleading terms in the action if the potential is nonzero.
\newsec{Supersymmetry Breaking}
In the eigenvalue description we are doing the quantum mechanics of $N$
discrete degrees of freedom.  In refs.~\refs{\Dav,\Dab}, the distribution of
$N-1$ eigenvalues
was used as a background to find the action for the $N$th eigenvalue.
In quantum mechanics, the presence of two ground states for the last eigenvalue
implies the existence of an instanton tunneling from one minimum to the other,
and the instanton effects give rise to a nonperturbative lifting of the ground
state energy.

In the field theory we have a number of candidates for the classical vacuum
field configuration in eq.~\grsttwo\ for varying values of $\alpha$, and in
eq.~\groundzero.  Are there instantons in this theory which connect any of
these putative ground states?  One may study this question by expanding the
action for the bosonic sector of the theory around the background field
configuration $\phi_0$ in \groundzero.  Since all the configurations described
by eq.~\grsttwo\ match $\phi_0$ to leading order in $1/N$ this background is a
useful way to study ground states and instantons.

The instanton of \Dab\ is described in our language by separating out a
$\delta$-function from $\phi$ as in \grsttwo\ where now the position of the
$\delta$-function is time dependent.  The leading order
Lagrangian for the bosonic part of the theory is:
\eqn\boslagr{
L = \half {(\partial^{-1} \dot{\phi})^2 \over \phi} - \half \left(W'(x)
       - {1\over N} \pint {dy \, \phi(y) \over x-y} \right)^2 \phi.
}
This Lagrangian may have instantons connecting different
ground states.  The instanton equation, satisfied by a minimum of the action
in Euclidean time, is:
\eqn\instanton{
\pm {(\partial^{-1} \dot{\phi}) \over \phi} =
        W'(x) - {1\over N} \pint{dy \, \phi(y) \over x-y},
}
and the instanton action is
\eqn\instact{
S_{\rm\scriptstyle inst} = \left. \int dx \phi(x) \left( W(x) - {1\over 2N}
       \pint dy \phi(y) \ln |x-y| \right) \right|^{t=+\infty}_{t=-\infty}.
}
In the approximation of neglecting back reaction, this gives rise to exactly
the
instanton of \Dab, but weighted by the factor $\alpha$.  To understand the
effect of this field configuration on supersymmetry breaking in the collective
field theory we have to understand how to
treat this parameter.

\newsec{Discussion}
We have thus found the supersymmetry collective field theory description of
the truncated Marinari Parisi model.  It
coincides to leading order in N with that discussed in \Jsup;
the derivation here makes clear the link to the surface interpretation
and also shows how the next order term in $N$
from the Jacobian appears.
The coefficient of the instanton at this level appears unfixed,
although it may be determined by subleading terms in the superpotential,
as happens when the potential vanishes.  The
larger instanton effects in the matrix model description of string
theory \shs\ than that expected in field theory has been associated with the
nonlocality and lack of translation invariance in the collective field
action\refs{\Jnp,\DJR}.

It would be interesting to make a connection to the
spacetime description of this theory.
There are many things known about the Marinari-Parisi
model, but its spacetime interpretation is not one of them.
Some observations were made in \Dab, mostly about the full supersymmetric
model.  Even the bosonic sector alone, corresponding to the
first $c=1$ higher multicritical theory, with an $x^3$ potential, has not been
identified.  The nonlocality in the bosonic sector of the
Marinari-Parisi models with higher order
superpotentials means they
do not naively correspond to $c=1$ multicritical points.
In the $d=1$ model,
the fluctuations around the static ground configuration
$\phi_0$ to leading order in $N$ describe a massless particle
related to the massless tachyon of the theory \refs{\poltachy,\DJ}.
For the cubic potential considered in this
paper, the `tachyon' is also massless, for higher order potentials
the nonlocality in \local\ gives the fluctuations
an effective mass at this order.  An effective mass appears in subleading order
in $1/N$ for all cases.
Writing $\phi(x,t) = \phi_0(x) + \partial_x\eta (x,t)$ and keeping only
second order in the fluctuations and leading order in $1/N$:
\eqn\fllag{
\eqalign{
L_{fluctuations} = \int dx\; \left[ {1\over 2N^2}\right. &
                     {(\dot{\eta})^2\over \phi_0}
- {\pi^2 \over 2 N^2} \phi_0(x) (\partial_x \eta(x))^2 \cr
 +& {1\over 2 N}  \pint dx\,dy\,{W'(x)-W'(y) \over x-y } \partial_x \eta(x)
\partial_y \eta(y)
 + {i\over 2}\int {dx\over \phi_0}(\bar{\psi} \dot{\psi} -
\bar{\dot{\psi}}\psi ) \cr
  - & {1\over N} W''(x){\psbar(x)\psi(x)\over\phi_0(x)}
- \left.
{1\over N^2} \psbar(x)\partial_x\pint dy {\psi(y)\over (x-y)} \right] \cr}
}
when $\phi_0(x) \ne 0$.

Much has been calculated for the bosonic higher multicritical local
potentials, generalizing from $d=1$, which can be compared with any
suggested spacetime interpretation.
The Hamiltonian can be rewritten
in terms of fermi sea momenta \pol, and again, most of the scattering takes
place at the boundaries of $x$ space.  An infinite number of symmetry
generators analogous to those found at $c=1$ (before double scaling)
\refs{\aj,\mpy}\
are present and presumably linked to the ground ring \witten\ structure
in the matter plus gravity theory.
The large order behavior in perturbation theory
and some scaling exponents\pg, the correlators of
the operators $M^n$ and the related analogues of the $c=1$ discrete
states \refs{\dg,\dan,\greg}\ have all been calculated, in part using the
Virasoro constraints \refs{\dg,\dan} present.

One could try to deduce directly how the modifications of the $-x^2$ potential
appear as modifications of $c=1$.
In the one matrix model, as $n$ in the potential $x^n$ increases,
the value of $c_{matter} \rightarrow - \infty$ for the matter sector.
For the two matrix model \twomat,
taking the same criticality in the potential $x^n$ for both matrices,
$c_{matter}$ increases as $n$ increases.
One suggestion \dan\ for these multicritical potentials,
based on their Wheeler--DeWitt equation,
is that the matter sector remains unchanged (leaving open the possibility of
altering the standard ghost-Liouville mixing as in \refs{\vip,\marte}).

The understanding of what matrix model instantons are in either the Liouville
or spacetime background picture would give clues to how they appear
in more general string backgrounds.
For instance, the effective operator for the $c=1$
instanton found in \Jnp\ should induce the phase shifts found in \gregpl.
Representing the string instanton
processes as effective operators in the theory defined around the
usual ground state would allow a better study of consequences of
these nonperturbative effects in string theory.

Acknowledgements:
We thank L. Brekke, S. Chaudhuri, A. Dabholkar, J. Lykken, E. Raiten,
and S. Shenker for discussions, J.C. also thanks M. Bawendi, F. David,
P. Ginsparg, A. Jevicki, V. Periwal and D. Spector.

Note added:  After we submitted this paper for publication, we learned
of the final version of \shahar, which uses a variable much like
the collective field in the $d=0$ matrix model.  An important difference
is that the analogue of the subleading term in the Jacobian is under more
control and so can be used more reliably.  It may be that the
results for instantons found there carry over naturally to the case
discussed in this paper.
\listrefs
\bye